\documentclass[graybox]{svmult}

\usepackage{cite}
\usepackage{type1cm}        
\usepackage{makeidx}         
\usepackage{graphicx}        
\usepackage{multicol}        
\usepackage[bottom]{footmisc}

\usepackage{newtxtext}       %
\usepackage{newtxmath}       

\makeindex             

\usepackage{hyphenat}

\newcommand{\nnl}{\nonumber\\}

\newcommand{\ZC}{\mathbb C}

\newcommand{\ZR}{\mathbb R}
\newcommand{\ZZ}{\mathbb Z}

\DeclareMathOperator{\zm}{\zeta}

\DeclareMathOperator{\Gt}{\tilde{\Gamma}}

\newcommand{\eqn}[1]{eq.~\eqref{#1}}

\DeclareMathOperator{\bC}{\mathbf{C}}

\newcommand{\El}{\tau}

\DeclareMathOperator{\Sel}{S}
\DeclareMathOperator{\SelE}{S^{\El}}
\DeclareMathOperator{\Selbld}{\mathbf{S}}
\DeclareMathOperator{\SelbldE}{\mathbf{S}^{\El}}
\DeclareMathOperator{\SelbldEw}{\mathbf{S}^{\El}_{\mathit{w}}}

\newcommand{\gd}[1]{g^{(#1)}}

\newcommand{\pd}{\partial}
\newcommand{\colvec}[1]{\begin{pmatrix}#1\end{pmatrix}}
\newcommand{\wmax}{{w_\text{max}}}
\def\oneloop{\text{1-loop}}
\newcommand{\ap}{\alpha'}
      
\newcommand{\CO}{\mathcal{O}}      

\newcommand{\half}{\tfrac{1}{2}}

\usepackage{xparse}

\ExplSyntaxOn
\NewDocumentCommand{\Gtarg}{m m m}
{
 \Gt\left(\begin{smallmatrix}
 \Gtarg_print:n {#1} \\
 \Gtarg_print:n {#2}
 \end{smallmatrix};#3\right)
}
\seq_new:N \l_Gtarg_list_seq
\cs_new_protected:Npn \Gtarg_print:n #1
{
  \seq_set_split:Nnn \l_Gtarg_list_seq { , } { #1 }
  \seq_use:Nn \l_Gtarg_list_seq { , & }
}
\ExplSyntaxOff

\ExplSyntaxOn
\NewDocumentCommand{\Gtargz}{m m}
{
 \Gt\left(\begin{smallmatrix}
 \Gtargz_print:n {#1} \\
 \Gtargz_print:n {#2}
 \end{smallmatrix};z\right)
}
\seq_new:N \l_Gtargz_list_seq
\cs_new_protected:Npn \Gtargz_print:n #1
{
  \seq_set_split:Nnn \l_Gtargz_list_seq { , } { #1 }
  \seq_use:Nn \l_Gtargz_list_seq { , & }
}
\ExplSyntaxOff

\ExplSyntaxOn
\NewDocumentCommand{\Gtargzt}{m m}
{
 \Gt\left(\begin{smallmatrix}
 \Gtargzt_print:n {#1} \\
 \Gtargzt_print:n {#2}
 \end{smallmatrix};z,\tau\right)
}
\seq_new:N \l_Gtargzt_list_seq
\cs_new_protected:Npn \Gtargzt_print:n #1
{
  \seq_set_split:Nnn \l_Gtargzt_list_seq { , } { #1 }
  \seq_use:Nn \l_Gtargzt_list_seq { , & }
}
\ExplSyntaxOff
\newcommand{\SI}[1]{\Sel[#1]}

\ExplSyntaxOn
\NewDocumentCommand{\SIE}{m m}
{
\SelE\!\Big[\begin{smallmatrix}
 \SI_print:n {#1} \\
 \SI_print:n {#2}
 \end{smallmatrix}\Big]
}
\seq_new:N \l_SI_list_seq
\cs_new_protected:Npn \SI_print:n #1
{
  \seq_set_split:Nnn \l_SI_list_seq { , } { #1 }
  \seq_use:Nn \l_SI_list_seq { , & }
}
\ExplSyntaxOff

\makeatletter
\providecommand*{\shuffle}{%
  \mathbin{\mathpalette\shuffle@{}}%
}
\newcommand*{\shuffle@}[2]{%
  \sbox0{$#1\vcenter{}$}%
  \kern .15\ht0 
  \rlap{\vrule height .25\ht0 depth 0pt width 2.5\ht0}%
  \raise.1\ht0\hbox to 2.5\ht0{%
    \vrule height 1.75\ht0 depth -.1\ht0 width .17\ht0 %
    \hfill
    \vrule height 1.75\ht0 depth -.1\ht0 width .17\ht0 %
    \hfill
    \vrule height 1.75\ht0 depth -.1\ht0 width .17\ht0 %
  }%
  \kern .15\ht0 
}
\makeatother

\newif\ifjbnote 
\jbnotetrue

\newif\ifaknote 
\aknotetrue

\begin{document}

\title*{A geometrical framework for amplitude recursions: bridging between trees and loops}
\author{Johannes Broedel and Andr\'e Kaderli}
\institute{Johannes Broedel \at Humboldt University, Unter den Linden 6, 10117 Berlin, 
\email{jbroedel@physik.hu-berlin.de}
\and André Kaderli \at Humboldt University, Unter den Linden 6, 10117 Berlin, \email{kaderlia@physik.hu-berlin.de}}
%
%
\maketitle
\begin{small}HU-Mathematik-2021-03, HU-EP-21/20\end{small}\\

\abstract*{
	Various methods for the recursive evaluation of scattering amplitudes in quantum field theory and string theory have been put forward during the last couple of years. In these proceedings we describe a geometrical framework, which is believed to be capable of treating many of these recursions in a unified way. 
Our recursive framework is based on manipulating iterated integrals on Riemann surfaces with boundaries. A geometric parameter appears as variable of a differential equation of KZ or KZB type. The parameter interpolates between two associated regularized boundary values, which contain iterated integrals closely related to scattering amplitudes defined on two different geometries.  
}

\abstract{
	Various methods for the recursive evaluation of scattering amplitudes in quantum field theory and string theory have been put forward during the last couple of years. In these proceedings we describe a geometrical framework, which is believed to be capable of treating many of these recursions in a unified way. 
Our recursive framework is based on manipulating iterated integrals on Riemann surfaces with boundaries. A geometric parameter appears as variable of a differential equation of KZ or KZB type. The parameter interpolates between two associated regularized boundary values, which contain iterated integrals closely related to scattering amplitudes defined on two different geometries.  
}

\section{Introduction}
\label{sec:introduction}
The calculation of scattering amplitudes in perturbative quantum field theories relies on the evaluation of Feynman integrals associated to Feynman graphs, which in turn are a combinatorial representation of Feynmans path integral formalism. A typical Feynman integral associated to an $\ell$-loop process reads
\begin{equation}
\label{eqn:Feynman1}
    (\mu^2)^{\nu-\frac{\ell D}{2}}\int\prod_{r=1}^\ell\frac{d^D k_r}{i \pi^{D/2}}\prod_{j=1}^n\frac{1}{(-{q_j}^2+m_j^2)^{\nu_j}},\quad\nu=\sum_{j=1}^n\nu_j\,.
\end{equation}
where $k_i$ are the loop momenta, $q_j$ and $m_j$ label the momenta and masses along the $n$ (internal and external) propagators. The quantity $D$ is the (spacetime) dimension and the integral shall usually be evaluated in four dimensions.

From the Feynman formalism, a multitude of different types of integrals can arise \cite{Goncharov:2010jf,Ablinger:2017bjx,Broedel:2017kkb}. A first step towards treating the integrals in a uniform way is to introduce Feynman parameters $x_i$, which amounts to a clever substitution of the momentum integrations in the Feynman integral \eqref{eqn:Feynman1}. Leaving out a constant prefactor, one obtains an integral of the following type:
\begin{equation}
    \int_{x_j\geq 0}\delta\big(1-\sum_{j=1}^n x_j\big)
    \Big(\prod_{j=1}^n dx_j 
    \Big)\,
    {{\cal I}(x_1,\ldots,x_n,D)}\,.
\end{equation}
Integration is constrained to a simplex by the condition $x_j\geq0$ and the $\delta$ distribution. Every simplex, however, can be parametrized iteratively. Consequently, each Feynman integral can be rewritten as a linear combination of iterated integrals. 

The type of iterated integral, on the other hand, is determined by the differential form $\cal I$, which shall to be integrated over. This differential form can have singularities in the Feynman parameters. The singularity structure can be explored by writing the integrand as 
\begin{equation}
	{\cal I}(x_1,\ldots,x_n)=\frac{\cal N}{\cal D}
\end{equation}
where the integrand becomes singular, whenever the denominator polynomial ${\cal D}$ has a zero\footnote{The polynomials ${\cal N}$ and ${\cal D}$ are very closely related to the Symanzik polyonomials.}. The zero locus of this polynomial defines an algebraic curve, which can be taken as starting point for the definition of suitable differentials incorporating the symmetries of the Feynman diagram. From those differentials, one can then build iterated integrals.  
Once the differential forms and associated iterated integrals are known, it is usually possible to write down a differential equation for a set of master integrals. The resulting (matrix) equation should hopefully be translatable into typical differential equations for a set of master integrals as used heavily in modern Feynman as well as string-theoretic calculations \cite{Kotikov:1990kg,Remiddi:1997ny,Gehrmann:1999as,Henn:2013pwa,Puhlfuerst:2015gta,Puhlfuerst:2015zqw}.  

While this mathematical account sounds very straightforward, it is peppered with practical difficulties: identification of suitable differential forms - that is, a cohomology - for a given algebraic curve is for example possible only for the simplest Feynman graphs. 

Therefore, in these proceedings, we will take the two simplest algebraic curves, Riemann surfaces with boundary of genus zero and genus one, as examples. The corresponding differential forms generate iterated integrals, which are polylogarithms (genus zero) or elliptic analogues thereof (genus one). Whereas actual Feynman integrals might imply more complicated differential forms, almost all final results turn out to be expressible in terms of these simple iterated integrals and special values thereof: multiple zeta values (MZVs) and elliptic multiple zeta values (eMZVs). 

The only structural ingredient we need to add is an extra parameter, with respect to which a differential equation governing the recursion is established. For Feynman integrals, this would be an additional Feynman parameter, while for string configuration-space integrals the parameter describes an additional vertex insertion point. 

So the recursive algorithms discussed in these proceedings are to be seen as prototypes for more complicated recursions. They are simple and thus mathematically very clean: they turn out to precisely describe recursions for amplitudes in open string theory at tree level (genus zero) and one-loop level (genus one).  

While Feynman integrals have to be regularized case-by-case (which is usually done with dimensional regularization), for the classes of iterated integrals considered in these proceedings a standard way of regularization is available. Thus, one will not have to find a suitable regularization for each Feynman integral separately, but can rather rely on a general regularization scheme for all integrals occurring.  

Pursuing this line of thoughts further, result for a scattering amplitude evaluated in the Feynman formalism is usually provided as loop expansion in the parameter $\ell$ and as expansion in the parameter $\epsilon$ of dimensional regularization. On the other hand, string scattering amplitudes defined as iterated integrals on Riemann surfaces are result in a genus expansion (parameter $g$) and an expansion in $\alpha'$, the inverse string tension. This suggestive correspondence might be substantiated by understanding ,,stringyness'' once again as a simple regulating mechanism, which after all is a very old idea. 

The whole subject is comparably involved algebraically, that is, there is a price to pay for the formalization. As a reward, the formalism is applicable to many different situations and is expected to lead to recursion relations for various types of iterated integrals and thus scattering amplitudes during the next couple of years. 

\section{Genus zero}
\label{sec:genuszero}

\subsection{Iterated integrals and multiple zeta values}
\label{ssec:Iteratedintegralsgenuszero}
Let us review the most straightforward implementation of polylogarithms on a genus-zero Riemann surface. Consider the one-form\footnote{For simplicity, we consider real integration paths here exclusively. More general quantities $a$, for example complex functions of complex parameters, will lead to the hyperlogarithms discussed in Erik 
Panzer's talk.}
\begin{equation}
	\label{eqn:diffforms}
	\omega_a=\frac{dx}{x-a},\quad x,a\in\ZR
\end{equation}
and define iterated integrals \cite{Goncharov:2001iea}
\begin{align}
	\label{eqn:Goncharithm}
	G(a_1,....,a_r;x)&=\int_0^x\frac{dt}{t-a_1}G(a_2,\ldots,a_r;t)=\int_0^z\omega_{a_r}\cdots\omega_{a_1}\,,\quad G(;z)=1\,,
\end{align}
where $a_1\neq z$ and $a_r\neq 0$. Given the iterated structure above, the integrals are subject to shuffle relations:
\begin{align}
\label{eqn:shuffle}
&G(a_1,a_2,\ldots,a_j; x)G(b_1,b_2,\ldots,b_k; x)\nnl
&\qquad=G\big((a_1,a_2,\ldots,a_j)\shuffle(b_1,b_2,\ldots,b_k); x\big)
\end{align}
and the differential forms in \eqn{eqn:diffforms} imply that different integrands in \eqn{eqn:Goncharithm} are related by partial fraction: 
\begin{equation}
\frac{1}{x_i-x_k}\frac{1}{x_j-x_k}=\frac{1}{x_i-x_j}\frac{1}{x_j-x_k}+\frac{1}{x_j-x_i}\frac{1}{x_i-x_k}\,.
\end{equation}

If the case $a_r=0$ was allowed, the integrals $G(a_1,\dots,a_r;x)$ would not be well-defined, since they would diverge due to the simple pole at the lower integration boundary. This can be regularized by the convention
\begin{align}
\label{eqn:Greg}
G(\underbrace{0,...,0}_{n};x)=\frac{\log^n x}{n!}\,,
\end{align}
such that integrals $G(a_1,\dots,a_r;x)$ associated to arbitrary labels $a_1,\dots, a_r\neq x$, including $a_r=0$, can be defined as follows: the shuffle identity \eqref{eqn:shuffle} is used to formally write and define the (a priori ill-defined) integral $G(a_1,\dots,a_r;x)$ as an expansion in the well-defined integrals \eqn{eqn:Goncharithm} and powers of logarithms $G(0,\dots,0;x)$. This regularization scheme is called \textit{shuffle regularization} and is compatible with (e.g.~it preserves) the shuffle product. The integrals $G(a_1,\dots,a_r;z)$ are called \textit{multiple polylogarithms} (MPLs).

For the purpose of exploring open-string amplitudes at tree level, it is sufficient to confine ourselves to labels $a_i\in\lbrace 0,1\rbrace$. Choosing $x=1$ leads to representation of \textit{multiple zeta values} (MZVs) in terms of iterated integrals:
\begin{align}
\label{eqn:translateMZV}
\zm(n_1,\ldots,n_r)&=(-1)^r G(\underbrace{0\ldots 01}_{n_r}\ldots\underbrace{0\ldots 01}_{n_1};1)\nnl
&=(-1)^r\int_0^1 \omega_1 \omega_0^{n_1-1}\omega_1\omega_0^{n_2-1}\ldots \omega_r \omega_0^{n_r-1}\nnl
&=\sum_{1\leq k_1<\dots<k_r}\frac{1}{k_1^{n_1}\dots k_r^{n_r}}\,,
\end{align}
for $n_r>1$. As before, this definition can be extended to arbitrary labels $n_1,\dots,n_r\geq 1$ and integrals $G(a_1,\dots,a_r;1)$ with $a_1=1$, respectively, by a shuffle regularization with the conventions
\begin{align}
	\zm(1)
	=-G(1,1)
	=0\quad\text{ as well as }\quad G(0;1)=\log 1 =0\,.
\end{align}
Multiple zeta values inherit shuffle relations from the polylogarithms; in addition there are the stuffle relations (best palpable in the sum representation in the last line of \eqref{eqn:translateMZV}). After considering all relations, a basis of MZVs at each conjectured transcendentality can be chosen, the mathematically most beautiful being the Hoffman basis~\cite{ZagierHofmann}.

\subsection{Selberg integrals and open-string configuration-space integrals}
Selberg integrals serve as generating series for the iterated integrals introduced in the previous subsection. At the same time they contain the configuration-space integrals appearing in open-string tree-level amplitudes. The full scattering amplitude in open superstring theories at tree level can be calculated as correlation function of vertex operators inserted on the boundary of a genus-zero Riemann surface. Upon evaluation, those correlators separate into a polarization part (which can be calculated straightforwardly) and the so-called configuration\hyp{}space integrals \cite{Mafra:2011nv,Mafra:2011nw,Broedel:2013tta}. The best known example is the four-point Veneziano amplitude \cite{Veneziano:1968yb}, which reads
\begin{equation}\label{eqn:ex4PtGenus0}
	\int_{0}^1 d x_3\, x_3^{s_{13}}(1-x_3)^{s_{23}} \frac{s_{13}}{x_3}=\frac{\Gamma(1+s_{13})\Gamma(1+s_{23})}{\Gamma(1+s_{13}+s_{23})}\,.
\end{equation}
The complex parameters 
\begin{equation}
	\label{eqn:Mandelstam}
	s_{i_1...i_r}=\ap(k_{i_1}+\ldots+k_{i_r})^2
\end{equation}
are Mandelstam variables built from the momenta of the external particles. In these proceedings, these variables shall be assumed to be chosen such that all integrals considered are convergent \cite{Mandelstam:1974fq,Brown:2019wna}. In contrast to the usual Mandelstam variables, a parameter $\alpha'$ is supplemented here, which serves as counting parameter and will be identified with the inverse string tension only later on. 

The $N$-point configuration\hyp{}space integrals in genus-zero open-string amplitudes are examples of Selberg integrals \cite{Selberg44}, which can be constructed as follows: consider the $(L{+}1)$-punctured Riemann sphere with fixed points
\begin{align}\label{eqn:fixedPunctures}
(x_1,x_2,x_{L+1})&=(0,1,\infty)\,.
\end{align}
Writing
\begin{align}
x_{ij}&=x_{i,j}=x_i-x_j\,,
\end{align}
the corresponding Selberg integrals are iteratively defined by
\begin{align}
	\label{eqn:SelbergzeroIntro}
	\SI{i_{k+1},\dots,i_L}(x_1,\dots,x_k)&=\int_0^{x_k}\frac{dx_{k+1}}{x_{k+1,i_{k+1}}}\SI{i_{k+2},\dots,i_L}(x_1,\dots,x_{k+1})\,,
\end{align}
and the empty Selberg integral (or Selberg seed) is defined as\footnote{We use the notation $\prod_{x_a\leq x_i<x_j\leq x_b}=\prod_{i,j\in \{1,2,\dots,L\}:\,x_a\leq x_i<x_j\leq x_b}$.}
\begin{equation}
	\label{eqn:SelbergzeroSeedIntro}
	\SI{}(x_1,\dots,x_L)
	=\prod_{0\leq x_i<x_j\leq 1}|x_{ij}|^{s_{ij}}\,. 
\end{equation}
Definition \eqref{eqn:SelbergzeroIntro} presumes that the so-called admissibility condition
\begin{equation}\label{eqn:admissibilityIntro}
	1\leq i_{p}< p\qquad\forall p\in\lbrace k+1,\ldots, L\rbrace
\end{equation}
is met. The integral in \eqn{eqn:SelbergzeroIntro} is referred to as of type $(k,L{+}1)$. It is, for fixed $s_{ij}$, defined on $\mathcal{M}_{0,k+1}$. Accordingly, these integrals form a basis of the twisted de Rham cohomology of the configuration space of $(L{+}1)$-punctured Riemann spheres with $k+1$ fixed punctures with respect to the pull-back of the connection $d+d\log \Sel$ \cite{aomoto1987}. Integrals with at least one label $i_p=1$ may be reduced to this basis using integration by parts and partial fractioning.

\subsection{Recursion for open-string amplitudes at genus zero}
\label{ssec:recursiongenuszero}

Aomoto \cite{aomoto1987} and Terasoma \cite{Terasoma} showed that Selberg integrals of type $(2,L)$ can be obtained algebraically from those of type $(2,L{-}1)$: one starts from a basis vector $\Selbld(x_3)$ for Selberg integrals of type $(3,L{+}1)$, which contain an auxiliary point $x_3$ in contrast to the integrals of type $(2,L)$ and $(2,L{-}1)$, respectively. Taking the derivative with respect to $x_3$ leads to an equation of Knizhnik--Zamolodchikov (KZ) type \cite{Knizhnik:1984nr}
\begin{equation}
\label{eqn:KZexampleIntro}
\frac{d}{d x_3}\Selbld(x_3) = \Big(\frac{e_0}{x_3}+\frac{e_1}{x_3-1}\Big)\Selbld(x_3)\,,
\end{equation}
where the (braid) matrices $e_0$ and $e_1$ have entries which are homogeneous polynomials of degree one in the parameters $s_{ij}$.  The regularized boundary values 
\begin{align}\label{eqn:regLimitGenusZero}
\bC_0 = \lim_{x_3 \rightarrow 0} x^{-e_0}\Selbld(x_3) \ , \ \ \ \bC_1 = \lim_{x_3\rightarrow 1} (1-x_3)^{-e_1} \Selbld(x_3)\,.
\end{align}
of the differential equation \eqref{eqn:KZexampleIntro} are Selberg integrals of type $(2,L-1)$ and $(2,L)$, respectively. They can be shown to be related by the Drinfeld associator \cite{Drinfeld:1989st,Drinfeld2}
\begin{align}
\label{eqn:genusZeroAssociatorEqIntro}
\bC_1&=\mathbf{\Phi}(e_0,e_1)\, \bC_0,
\end{align}
which is the generating series of multiple zeta values~\cite{LeMura},
\begin{align}
\mathbf{\Phi}(e_0,e_1)&=\sum_{w\geq 0}\sum_{k_1,\dots,k_w \geq 1}e_0^{k_w-1}e_1\dots e_0^{k_2-1}e_1e_0^{k_1-1}e_1\zeta(k_1,k_2,\dots,k_w)\nnl
		      &=1-\zeta(2) [e_0,e_1]-\zeta(3)\left([e_0,[e_0,e_1]]-[[e_0,e_1],e_1]\right)+\dots\,.
\label{eqn:leadingap}
\end{align}
What makes this construction useful for physicists is the fact that the $(N{-}1)$-point and the $N$-point configuration\hyp{}space integrals at genus zero can be identified (upon proper assignment of the Mandelstam variables) as linear combinations of the components of $\bC_0$ and $\bC_1$ respectively, where $N{=}L$. This relationship has been used to derive a recursive construction for all configuration\hyp{}space integrals on genus zero: it provides an analogue of the Parke--Taylor formula \cite{Parke:1986gb} for string theory \cite{Broedel:2013aza,Mafra:2016mcc}. The precise relation of the above formalism to open-string configuration\hyp{}space integrals at genus zero has been discussed thoroughly in ref.~\cite{AKtbp}.

\section{Genus one}
\label{sec:genusone}

\subsection{Iterated integrals at genus one and elliptic multiple zeta values}

While there is a large collection of literature \cite{Levin, BrownLev, LevinRacinet} on how to define homotopy-invariant integrals on an elliptic curve (or torus), we are going to focus here on what is one of the simplest approaches and simultaneously the best fit for a canonical generalization of the iterated integrals on a genus-zero surface introduced in \eqn{eqn:Goncharithm}. We will parametrize the elliptic curve by the modular parameter $\tau$ (or its exponentiated version $q=\exp(2\pi i\tau)$) and name the red and blue boundaries of the fundamental domain $A$- and $B$-cycle respectively.
\begin{figure}
	\includegraphics[width=0.9\textwidth]{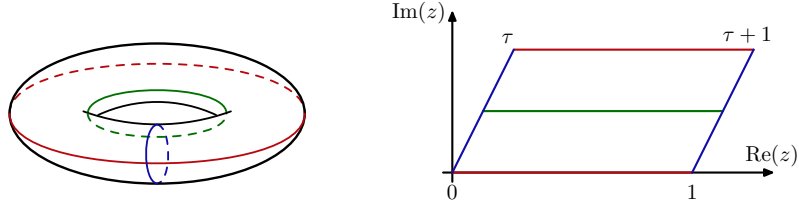}
	\caption{The torus and its fundamental domain. The ratio of the lengths $\omega_B$ and $\omega_A$ of the $B$- and $A$-cycle respectively yields the modular parameter: $\tau=\omega_B/\omega_A$.}
	\label{fig:fundamentaldomain}
\end{figure}
An (infinite) set of differential forms on the elliptic curve can be defined starting from the Kronecker series 
\begin{equation}
F(z,\eta,\tau)=\frac{\theta_1'(0,\tau)\theta_1(z+\eta,\tau)}{\theta_1(z,\tau)\,\theta_1(\eta,\tau)}\,,
\label{eqn:EK}
\end{equation}
where $\theta_1$ is the odd Jacobi function and the tick denotes a derivative with respect to the first argument. The Kronecker series is symmetric in $z$ and $\eta$, but only quasiperiodic in the variable $z$:
\begin{equation}
\label{eqn:quasiperiodic}
F(z+1,\eta,\tau)=F(z,\eta,\tau),\hspace{2em}F(z+\tau,\eta,\tau)=e^{-2\pi i \eta}F(z,\eta,\tau)\,.
\end{equation}
In addition, Fay's trisecant equation \cite{mumford1984tata} implies the Fay identity
\begin{align}
\label{eqn:FayF}
&F(z_1,\eta_1,\tau)F(z_2,\eta_2,\tau)\nnl
&\quad =F(z_1,\eta_1+\eta_2,\tau)F(z_2-z_1,\eta_2,\tau)+F(z_2,\eta_1+\eta_2,\tau)F(z_1-z_2,\eta_1,\tau) \ .
\end{align}
Expanding the Kronecker form in the second argument, one obtains an infinite set of differential forms
\begin{align}
	\label{eqn:Ftog}
	\eta F(z,\eta,\tau)dz &= \sum_{n=0}^{\infty}g^{(n)}(z,\tau)\,\eta^{n} dz, 
\end{align}
satisfying 
\begin{equation}
	\label{eqn:gparity}
\gd{n}(-z,\tau)dz=(-1)^{n}\gd{n}(z,\tau)dz.
\end{equation}
They are only quasi-periodic
\begin{subequations}
\begin{align}
	\gd{n}(z+1,\tau)&=\gd{n}(z,\tau)\\
	\gd{1}(z+\tau,\tau)&=\gd{1}(z,\tau)-2\pi i\\
	\gd{2}(z+\tau,\tau)&=\gd{2}(z,\tau)-2\pi i\gd{1}(z,\tau)-\half (2\pi i)^2\\
			   &\vdots\nonumber
\end{align}
\end{subequations}
and satisfy the (expanded) form of Fay relations:
\begin{align}
	g^{(n_1)}(t-z,\tau) g^{(n_2)}(t,\tau) &=  - (-1)^{n_1} g^{(n_1+n_2)}(z,\tau) \nnl
					      &\quad+\sum_{j=0}^{n_2} \binom{ n_1 - 1 + j}{j} g^{(n_2-j)}(z,\tau) g^{(n_1+j)}(t-z,\tau)\nnl
					      &\quad+\sum_{j=0}^{n_1} \binom{n_2-1+j}{j} (-1)^{n_1+j} g^{(n_1-j)}(z,\tau) g^{(n_2+j)}(t,\tau).
	\label{eqn:gFay}
\end{align}
Taking the differential forms $\gd{n}(z,\tau)$ as starting point, one defines the following iterated integrals:
\begin{equation}
\label{eqn:defGt}
\Gtargzt{n_1,n_2,\dots,n_r}{a_1,a_2,\dots,a_r}=
\int_0^zdz'\, g^{(n_1)}(z'-a_1,\tau)\Gtargzt{n_2,\dots,n_r}{a_2,\dots,a_r}\,,
\end{equation}
for $(n_1,a_1)\neq (1,z)$ and $(n_r,a_r)\neq(1,0)$, which naturally obey shuffle relations
\begin{align}
	\label{eqn:Gammashuffle}
	&\Gt(A_1,A_2,\ldots,A_j; z,\tau)\Gt(B_1,B_2,\ldots,B_k; z,\tau)\nnl
	&\qquad=\Gt\big((A_1,A_2,\ldots,A_j)\shuffle(B_1,B_2,\ldots,B_k); z,\tau\big)
\end{align}
in terms of the combined letters $A_i=\begin{smallmatrix}n_i\\a_i\end{smallmatrix}$.

The function $\gd{1}$ has a simple pole at zero: it thus qualifies as genus-one generalization of $\frac{1}{z}$ at genus zero. The integral over $g^{(1)}$ will be of particular interest below: it is the ($A$-cycle) generalization of the natural logarithm. Similar to the prescription in \eqn{eqn:Greg}, the integral $\Gtargzt{1}{0}$ is a priori not well-defined and requires regularization because it exhibits an endpoint divergence at the lower integration boundary. Throughout the article, we are going to employ \textit{tangential basepoint regularization} \cite{Deligne89,Brown:ICM14}. In short, this amounts to subtracting the endpoint divergence by defining\footnote{The limit $\epsilon\to 0$ is assumed to be taken within the unit interval.}
\begin{align}\label{eMPL:DefReg}
\Gt(\begin{smallmatrix}1\\0 \end{smallmatrix}; z,\tau)
&=\lim_{\epsilon\rightarrow 0}\int_{\epsilon}^z dz\, g^{(1)}(z,\tau) +\log(1-e^{2\pi i \epsilon})\nnl
&=\log(1-e^{2\pi i z})-\pi i z+4\pi \sum_{k,l>0}\frac{1}{2\pi k}\left(1-\cos(2\pi k z)\right)q^{kl}\,.
\end{align}
In particular, when placing the branch cut of the logarithm such that $\log(-1)=\pi i$, one finds the following asymptotic behavior for $z\to 0$ 
\begin{equation}\label{eqn:asympGammaz0}
\Gt(\begin{smallmatrix}1\\0 \end{smallmatrix}; z,\tau)\sim \log(-2\pi i z)
\end{equation}
and $z\to 1$
\begin{equation}\label{eqn:asympGammaz1}
\Gt(\begin{smallmatrix}1\\0 \end{smallmatrix}; z,\tau)\sim \log(-2\pi i (1-z)) \,.
\end{equation}
The remaining integrals $\Gtargzt{n_1,n_2,\dots,n_r}{a_1,a_2,\dots,a_r}$ with $(n_r,a_r)=(1,0)$ are then defined by shuffle regularization, similar to the genus-zero integrals from \eqn{eqn:Goncharithm}: they are defined by the well-defined iterated integrals from \eqn{eqn:defGt}, the shuffle identity \eqref{eqn:Gammashuffle} and the regularized integral $\Gt(\begin{smallmatrix}1\\0 \end{smallmatrix}; z,\tau)$. This regularization procedure is compatible with the shuffle product, i.e.~an algebra homomorphism. 
For the remainder of those proceedings, the $\tau$-dependence will be mostly kept implicit for all integration kernels $g^{(n)}$ and all iterated elliptic integrals $\tilde{\Gamma}$. The latter are called \textit{elliptic multiple polylogarithms} (eMPLs).

\subsection{Elliptic multiple zeta values}

In the same way, as multiple zeta values can be represented as values of a special class of MPLs at one, so-called $A$-cycle \textit{elliptic multiple zeta values} (eMZVs) \cite{Enriquez:Emzv,Matthes:Thesis,Broedel:2014vla} are defined as values of regularized eMPLs at one: 
\begin{align}
	\label{eqn:defAcyceMZV}
\omega(n_1,n_2,\ldots,n_r)  
%
&= \Gtarg{n_r ,\ldots ,n_1}{0 ,\ldots ,0}{1}\,,\quad n_1\neq 1.
\end{align}
In order to extend the definition \eqref{eqn:defAcyceMZV} to the cases $n_1=1$, eMZVs need to be further regularized in a shuffle-compatible way. While a thorough discussion of the regularization procedure starting from the regulated integral in \eqn{eMPL:DefReg} can be found in ref.~\cite{Broedel:2019gba}, regularization of $A$-cycle eMZVs practically amounts to defining
\begin{equation}
  \omega(1)=0
\end{equation}
and using shuffle relations (inherited from \eqn{eqn:Gammashuffle})
\begin{equation}
	\label{eqn:Acycshuffle}
\omega(n_1,n_2,\ldots,n_r) \omega(k_1,k_2,\ldots,k_s) = \omega\big( (n_1,n_2,\ldots,n_r) \shuffle (k_1,k_2,\ldots,k_s) 
\big) \ ,
\end{equation}
to identify and isolate all those contributions. Furthermore, \eqn{eqn:gparity} implies 
\begin{equation}
	\label{eqn:cycparity}
\omega(n_1,n_{2} ,\ldots ,n_{r-1} ,n_r)=
     (-1)^{n_1+n_2+\ldots+n_r}\omega(n_r,n_{r-1} ,\ldots ,n_2 ,n_1)\,.
\end{equation}
The two types of relations above by far do not exhaust all relations between elliptic multiple zeta values; in particular does the Fay identity \eqn{eqn:gFay} imply many more relations. A thorough discussion can be found in ref.~\cite{Broedel:2015hia} and a list of relations on the associated website~\cite{eMZVWebsite}.

\subsection{Generalized Selberg integrals at genus one}
In order to investigate a genus-one analogue of the genus-zero recursive construction in section~\ref{sec:genuszero}, we need a suitable analogue of genus-zero Selberg integrals \eqref{eqn:SelbergzeroIntro}: let $L\geq 2$, $0=z_1<z_L<...<z_2<1$ and $\tau$ the modular parameter of the torus $\ZC/(\ZZ+\tau\ZZ)$. Let the empty genus-one Selberg integral (or genus-one Selberg seed) be
\begin{equation}
\label{eqn:SelbergSeedEIntro}
\SelE=\SIE{}{}(z_1,\dots,z_L)=\prod_{0=z_1\leq z_i< z_j\leq z_2}\exp\left(s_{ij}\Gt_{ji}\right)\,,
\end{equation}
where $\Gt_{ji}=\Gt(\begin{smallmatrix}1\\0 \end{smallmatrix}; z_j-z_i,\tau)$. Genus-one Selberg integrals of weight $w=\sum_{i=k+1}^L n_i$ and type $(k,L)$ are then defined recursively by
\begin{align}
\label{eqn:SelbergIntro}
&\SIE{n_{k+1},\dots,n_L}{i_{k+1},\dots,i_L}(z_1,\dots,z_k)\nnl
&\phantom{bbbb}=\int_0^{z_k}dz_{k+1}\, g^{(n_{k+1})}_{k+1,i_{k+1}}\SIE{n_{k+2},\dots,n_L}{i_{k+2},\dots,i_L}(z_1,\dots,z_{k+1})\,.
\end{align}
where we use the shorthand notation
\begin{equation}
	\gd{n}_{ij}=\gd{n}_{i,j}=\gd{n}(z_i-z_j,\tau)\,.
\end{equation}
Moreover, the so-called admissibility condition $1\leq i_p<p$ is required for all $p\in \{k+1,\dots, L\}$, which is the genus-one analogue of \eqn{eqn:admissibilityIntro}.

To build a recursion following the structure of the genus-zero recursion reviewed in subsection~\ref{ssec:recursiongenuszero}, we need to find a suitable class of genus-one Selberg integrals: to achieve this, we fix the symmetries of the torus by $z_1=0$, supplement one unintegrated auxiliary point $z_2$, such that $k=2$ punctures are fixed and integrate over the remaining $L-2$ punctures, but keep the number $L$ of insertion points variable. The resulting class of genus-one Selberg integrals reads
\begin{align}
\label{eqn:relevantGenusOneSelberg}
\SIE{n_{3},\dots,n_L}{i_{3},\dots,i_L}(z_1=0,z_2)
&=\int\limits_{0=z_1<z_L<z_{L-1}<\dots<z_2}\prod_{i=3}^L dz_i\, \Sel^\El \prod_{k=3}^L g^{(n_k)}_{k,i_k}\,.
\end{align}

Again, we would like to identify a basis in the above class of integrals with respect to integration by parts and partial fractioning. While there was only one type of differential form in the genus-zero situation (which one could have assigned weight one), we have an infinite number here: all combinations of $n_3,...,n_L$ can appear and for each of those combinations (almost) all admissible values can occur. This combinatorial problem can be solved \cite{Broedel:2019gba} and we collect all basis elements in a vector $\SelbldEw(z_2)$ of definite weight $w$ and combine all those vectors into an infinitely large vector: 
\begin{equation}
	\label{eqn:Selz2}
	\SelbldE(z_2)=\colvec{
		\Selbld^\El_0(z_2) \\
		\Selbld^\El_1(z_2) \\
		\Selbld^\El_2(z_2)\\
		\vdots
	}\,.
\end{equation}
The resulting vector $\SelbldE(z_2)$ is the analogue if the genus-zero Selberg vector $\Selbld(x_3)$, which satisfies the KZ \eqn{eqn:KZexampleIntro}.

\subsection{Selberg recursion at genus one}

In this subsection, it will be argued that the derivative of the vector $\SelbldE(z_2)$ defined in \eqn{eqn:Selz2} with respect to the auxiliary point $z_2$ can be written in the form 
\begin{align}
\label{eqn:KZBz2Intro}
\frac{\pd}{\pd z_{2}} \SelbldE(z_2)
&=\sum_{n\geq 0}g^{(n)}_{21}x^{(n)} \SelbldE(z_2)\,,
\end{align}
where the non-vanishing entries of the matrices $x^{(n)}$ turn out to be homogeneous polynomials of degree one in the parameters $s_{ij}$. The resulting system is of elliptic KZB-type, whose solution will be described below.

Proving the above statement is elaborate and is spelt out in detail in ref.~\cite{Broedel:2019gba}. The proof is constructive and relies on formal explicit evaluation of the derivative for each entry of the vector $\SelbldE(z_2)$. Performing the derivative on the Selberg seed will bring down various terms of the form
\begin{align}
\label{eqn:derivativeSE}
\frac{\pd}{\pd z_i}\SelE&=\sum_{k\neq i}s_{ik}\,g^{(1)}_{ik}\SelE\,,
\end{align}
whereas all other derivatives can be rewritten using integration by parts as to act on the Selberg seed exclusively. Thus one is left with Selberg integrals of definite length containing products of functions $g_{ij}^{(n)}$. Organizing these products in so-called \textit{chains} (e.g.~$g_{ij}^{(n_1)}g_{jk}^{(n_2)}g_{kl}^{(n_3)}$) allows to translate the application of Fay identities into graphical operations. Employing the (graphical analogue of) Fay identities algorithmically, one can show that in each of those integrands a factor $g_{21}^{(n)}$ can be isolated. Pulling this factor out of the integral (as neither the point $z_1$ or $z_2$ are integrated over) renders the remaining integral a basis integral, that is, a component of the original vector $\SelbldE(z_2)$. Accordingly, the Mandelstam variables arising from \eqn{eqn:derivativeSE} can then be collected in the matrices $x^{(n)}$, yielding the closed system in \eqn{eqn:KZBz2Intro}~\cite{Broedel:2019gba}.  

\begin{figure}[h!]
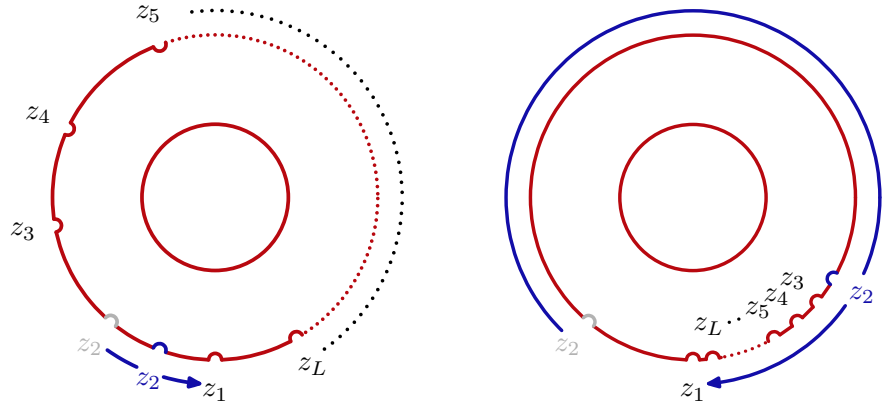

	\includegraphics[width=0.45\textwidth]{geometricamplitudes-genusonez2to1.mps}
	\hspace{0.1\textwidth}
	\includegraphics[width=0.43\textwidth]{geometricamplitudes-genusonez2to0.mps}
	\caption{Limits of the auxiliary point $z_2$ correspond to approaching the point $z_1=0\equiv1$ -- the origin of the fundamental domain -- along the real line from the left and from the right. While the limit $z_2 \to 1$ describes a smooth merging of $z_2$ with the point 1, in the limit $z_2 \to 0$ the other points $z_i$ are squeezed in the vanishing interval close to zero.}
	\label{fig:limits}
\end{figure}

What remains, is to solve \eqn{eqn:KZBz2Intro}. In the same way as this has been done for the KZ-system in subsection \ref{ssec:recursiongenuszero}, one can solve the system by considering regularized boundary values, which are related by the elliptic KZB associator \cite{Bernard:1987df,Bernard:1988yv}.
Regularized boundary values for the KZB system in \eqn{eqn:KZBz2Intro} are defined as
\begin{align}
	\label{eqn:bdryvalues}
	\bC_1^\El=\lim_{z_2\rightarrow 1}(-2 \pi i (1{-}z_2))^{-x^{(1)}}\SelbldE(z_2)\text{ and }
\bC_0^\El=\lim_{z_2\rightarrow 0}(-2 \pi i z_2)^{-x^{(1)}}\SelbldE(z_2)\,.
\end{align}
These two boundary values are related (see e.g.~\cite{Broedel:2019gba}) by the elliptic KZB associator $\Phi(x^{(0)},x^{(1)},x^{(2)},...)$ via
\begin{equation}\label{eqn:assocEqIntro}
\bC_1^\El=\Phi(x^{(0)},x^{(1)},x^{(2)},...)\bC_0^\El\,,
\end{equation}
whereas the KZB associator is -- in analogy to the KZ associator in \eqn{eqn:leadingap} -- a generating series for $A$-cycle eMZVs:
\begin{align}
	\Phi^\El&=1 + x^{(0)} - 2 \zeta(2) x^{(2)} \nnl
		&\quad+ \frac{1}{2} x^{(0)}x^{(0)} - (x^{(0)}x^{(1)}-x^{(1)}x^{(0)}) \omega(0, 1;\tau)  - \zeta(2) (x^{(0)}x^{(2)} + x^{(2)}x^{(0)}) \nnl
		&\quad+ \big(x^{(1)}x^{(2)}-x^{(2)}x^{(1)}\big)\big(\omega(0,3;\tau)-2\zeta(2)\omega(0,1;\tau)\big)
		+ 5 \zeta(4) x^{(2)}x^{(2)} + \cdots
		    \label{eqn:KZBass}
\end{align}
Equation \eqref{eqn:assocEqIntro} is the main tool in the recursive construction at genus one. What remains to be done before it can be applied, is the investigation of the boundary values $\bC_0^\El$ and $\bC_1^\El$ pictured in figure~\ref{fig:limits}.
Careful evaluation of the boundary values is beyond the scope of these proceedings, but is performed in detail in \cite{Broedel:2019gba}. The analysis relies on evaluating the matrix exponential in \eqn{eqn:bdryvalues}, and thus requires consideration of eigenspaces and eigenvalues of the braid matrix $x^{(1)}$. Furthermore, when calculating results for Selberg integrals using \eqn{eqn:assocEqIntro}, one has to limit the size of the system: the infinitely long vector $\SelbldE(z_2)$ has to be cut to finite length, i.e.~one needs to consider entries up to a certain weight $\wmax$ only. The maximal weight, in turn depends on the order in $\alpha'$ the expansion of the Selberg integral shall be calculated. Dependencies and the process of cutting the system to finite size is again carefully examined in ref.~\cite{Broedel:2019gba}, but will be used in the example in subsection \ref{ssec:example} below. 

Similarly to the previous genus-zero section, the regularized boundary value $\bC_1^\El$ can be shown to contain $(L{-}1)$-point configuration\hyp{}space integrals at genus one \cite{Green:1982sw,Dolan:2007eh,Broedel:2014vla,Mafra:2019xms} whereas $\bC_0^\El$ contains $(L{+}1)$-point configuration\hyp{}space integrals at genus zero. Accordingly, the $N$-point configuration\hyp{}space integrals appearing in open-string amplitudes at genus one can be calculated from the $(N{+}2)$-point integrals at genus zero via \eqn{eqn:assocEqIntro}, with $N{=}L{-}1$. This is going to be exemplified in the next subsection. 

\subsection{Recursive evaluation of two-point open-string integrals at genus one}
\label{ssec:example}

The successful concept for the calculation of genus-zero string-integrals from Selberg integrals, will be extended to genus one here.  
One-loop open-string amplitudes are calculated on a genus-one Riemann surface with boundary. Setting up the string correlation function, the problem can be divided in a polarization part and configuration\hyp{}space integrals. Omitting the (rather straightforward) polarization part, we will furthermore limit our attention to those configuration\hyp{}space integrals where points are inserted on one boundary only. 

Instead of developing the full theory here, let us present the easiest nontrivial example: the two-point case, which we would like to calculate to second order in $\ap$. The two-loop correction yields non-trivial results only, if the Mandelstam variables $s_{ij}$ are treated as independent parameters of the integrals, which do not satisfy any constraints like momentum conservation.

The two-point configuration\hyp{}space integral reads \cite{Mafra:2019xms}
\begin{equation}\label{eqn:2loopCorrection}
S^\oneloop_{2\text{-point}}(\tilde{s}_{13})
=\int_0^1 dz_3 \exp\left(\tilde{s}_{13} \Gt_{31}\right)=\sum_{n\geq 0}\tilde{s}_{13}^n\,\omega(\underbrace{1,\dots,1}_{n},0)\,,
\end{equation}
where the Mandelstam variable $\tilde{s}_{13}$ is associated to the loop momentum. Requiring two vertex insertion, the appropriate genus-one Selberg integral with an extra insertion point $z_2$ has length $L=3$ and the insertion points on the cylinder boundary are ordered as 
\begin{equation}
  0=z_1<z_3<z_2<1\equiv z_1\text{ mod }\ZZ\,.
\end{equation}
In the limit $z_2\to 1$, the punctures $z_2$ and $z_1$ merge, leaving
us with two punctures for the one-loop string corrections. Accordingly, we
are advised to consider the integrals
\begin{align}
	\label{eqn:twopointintegrals}
	\SIE{n_3}{i_3}(0,z_2)&=\int_0^{z_2} dz_3 \Sel^\El g^{(n_3)}_{3 i_3}\,,\quad 1\leq i_3 <3\,,\nnl
	\Sel^\El&=\exp\left(s_{13}\Gt_{31}+s_{12}\Gt_{21}+s_{23}\Gt_{23}\right)\,.
\end{align}

In the same way as the components of the vector $\SelbldE(z_2)$ are ordered by weight, so are the vectors $\bC_1^\El$ and $\bC_0^\El$. While the two-point one-loop correction is contained in the weight-zero entry, the tree-level correction can be found in the weight-one component. Sorting out the details, the goal of calculating up to second order in $\alpha'$ implies maximal weight two in the KZB system.

At this point, we would like to refer the reader to ref.~\cite{Broedel:2019gba} for (actually a lot of) careful derivation and write down the explicit two-point realization of \eqn{eqn:assocEqIntro} right away:
\begin{equation}\label{eqn:assEqn2Point}
\begin{pmatrix}
S^\oneloop_{2\text{-point}}(\tilde s_{13})
\\
\ast\\
\ast\\
\ast
\end{pmatrix}+{\cal O}\Big((\ap)^{3}\Big)=\Phi_3^\El(x^{(n)}_{\leq 2})\, \begin{pmatrix}0\\
\frac{1}{s_{13}}\frac{\Gamma(1+s_{13})\Gamma(1+s_{23})}{\Gamma(1+s_{13}+s_{23})}\\
0\\
0
\end{pmatrix}\,,
\end{equation}
where $\tilde{s}_{13}=s_{13}+s_{23}$ and only a finite part of the associator (cf.~\eqn{eqn:KZBass}) has to be determined. As indicated by the subscript of the associator in \eqn{eqn:assEqn2Point}, to calculate the one-loop configuration-space integral up to the second order in $\ap$, products of at most three (cut) matrices $x^{(n)}$ have to be included, which are given by:
\begin{equation}\label{eqn:2ptExamplex01}
x^{(0)}_{\leq 2}=\begin{pmatrix}
0&s_{13}&0&0\\
0&0&-s_{23}&-s_{23}\\
0&0&0&0\\
0&0&0&0
\end{pmatrix},\,\, x^{(1)}_{\leq 2}=\begin{pmatrix}
s_{12}&0&0&0\\
0&s_{123}&0&0\\
0&0&s_{12}+s_{23}&-s_{23}\\
0&0&-s_{13}&s_{12}+s_{13}
\end{pmatrix}
\end{equation}
and
\begin{align}\label{eqn:2ptExamplex2}
x^{(2)}_{\leq 2}&=\begin{pmatrix}
0&0&0&0\\
-s_{23}&0&0&0\\
0&s_{13}&0&0
\\
0&s_{13}&0&0
\end{pmatrix}\,.
\end{align}
Putting everything together, the relevant subpart of the matrix \eqn{eqn:assEqn2Point} reads
\begin{align}\label{eqn:assEqn2PointRelevant}
&S^\oneloop_{2\text{-point}}(\tilde{s}_{13})+{\cal O}\Big((\ap)^{3}\Big)\nnl
&\qquad\qquad=\Phi_3^\El\Big(x^{(n)}_{\leq 2}\Big)_{0,1} \frac{1}{s_{13}}\frac{\Gamma(1+s_{13})\Gamma(1+s_{23})}{\Gamma(1+s_{13}+s_{23})}\nnl
&\qquad\qquad=1+(s_{13}+s_{23})\omega(1,0)+(s_{13}+s_{23})^2\omega(1,1,0)+\CO\Big((\ap)^3\Big)\,,
\end{align}
where the subindex on the associator $\Phi$ specifies the appropriate matrix component. 
Nicely enough, this reproduces indeed the two-point one-loop string correction $S^\oneloop_{2\text{-point}}(\tilde{s}_{13})$ given in
\eqn{eqn:2loopCorrection} with the effective Mandelstam variable
$\tilde{s}_{13}=s_{13}+s_{23}$ up to second order in $\ap$.

\subsection{Geometric interpretation}
\label{ssec:geometric}
What is the geometric meaning of the two limits $z_2\to 0$ and $z_2\to 1$ in the genus-one case? The latter limit has an easy explanation: the merging of the point $z_2$ with the point $0\equiv 1$ happens in exactly the right way as to yield a finite result from two competing processes: the regularization of the boundary value and the behavior of the function $\Gt$ for $z_2$ close to one. The resulting geometry is just the same as one has been starting with: just a point less. 
\begin{figure}[h]
	\includegraphics[width=\textwidth]{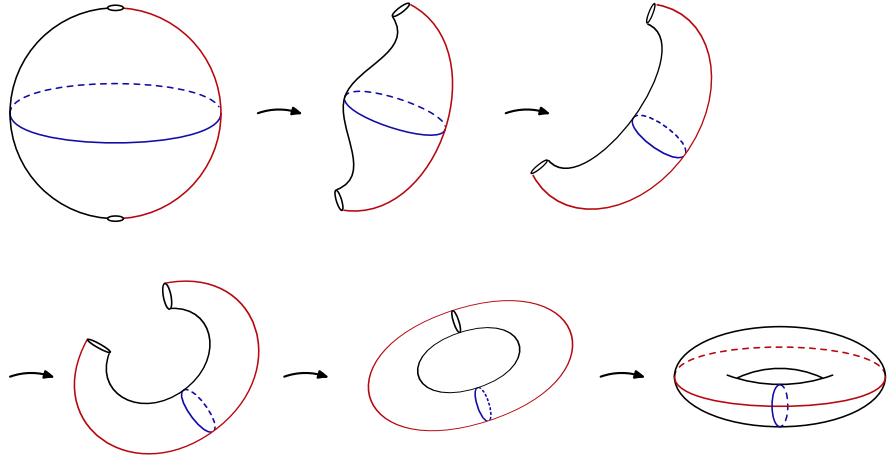}
	\caption{Step by step morphing of the Riemann sphere to the torus by joining infinitesimal circles at north- and south pole. The reverse process is modeled by the limit $\tau\to i\infty$. }
	\label{fig:morphing}
\end{figure}
More involved is the other limit: when $z_2$ tends to zero, all other insertion points $z_3$ to $z_L$ are squeezed in the (infinitesimal) interval $(0,z_2)$. Effectively, this limit amounts to \textit{shortening} the elliptic $A$-cycle: this implies that the modular parameter $\tau$, which is the ratio of the lengths of $B$- and $A$-cycle becomes very large (cf.~fig.~\ref{fig:fundamentaldomain}). Simultaneously, as can be justified by an integral transformation \cite{Broedel:2019gba}, from the perspective of the Riemann sphere the red line (which used to be the $A$-cycle before) becomes infinitely long, resulting in half a great circle (the positive real axis) on the Riemann sphere (see figure~\ref{fig:morphing}).

\section{General framework and outlook}

\subsection{What does it need for a general recursion? }
The two recursive algorithms reviewed above have several structural commonalities, however, are rather different when considering limits of the respective associated differential equations. 

In both formalisms, an algebraic variety is taken as starting point and an integrable connection is associated. This connection is built from differential forms with at most simple poles, thus leading to logarithmic singularities after iterated integration. The differential forms incorporate the periodicities/cycles of the algebraic variety in question. Symmetries of the variety, for example the choice of origin, are implemented by fixing a couple of positions in the Selberg integral, which simultaneously singles out a canonical path for the integration using homotopy invariance.  

In a next step, a differential equation with respect to an auxiliary point shall be established. For simplicity, let us assume the iterated integration to happen in the interval $[0,1]$, which is divided by several insertion points. (This is the case for both recursions discussed above.) In both scenarios, the auxiliary point is placed between the largest insertion point and one. 

Once the auxiliary point, which is a parameter not to be integrated over, has been supplemented in a set of integrals, one can now identify a basis set of integrals and determine the derivative. It is not yet clear, what a necessary or sufficient condition for closure of this system of differential equations is: in the two scenarios above we have just been lucky (or standard enough).  

To this end, one shall consider the boundary values. In an intricate interplay between regularization of the integrals, the regularization of the boundary values one can relate iterated integrals without auxiliary point featuring different numbers of insertion points and thus integrations.   

While all of the above considerations have been fairly general, the geometric interpretation finally depends on the particular surface in question, on particular on its cycles. The geometric picture incorporated by taking the two limits of the differential equation in the genus-one case are discussed in subsection~\ref{ssec:geometric} above. 

While there are several further examples, where a similar approach has been successful, let us here mention the recent calculation of the maximal cut of multiloop banana amplitudes in refs.~\cite{Klemm:2019dbm,Bonisch:2020qmm}. The ingredients are very similar: there is a (slightly more complicated) algebraic variety: a Calabi-Yau manifold, a Picard-Fuchs type differential equation (this time without auxiliary point), a basis set of integrals determined from the cohomology of the Calabi-Yau manifold. As turns out, the ideal of this Picard-Fuchs is a Gelfand--Kapranov--Zelevinsky(GKZ)-system, which delivers the desired result.\medskip 

A final remark is in place here: Feynman integrals are associated to graphs with edges, while string amplitudes are expressed as correlation functions on two-dimensional worldsheets. Considering the results, however, there is always a way to replace the Feynman expression with a set of iterated integrals naturally defined on a Riemann surface. 
Even more: when taking (dimensional) regularization into account, the result of calculating a particular scattering process using the Feynman formalism will be a double expansion: the topological expansion in the number of loops $\ell$ and the expansion in the parameter $\epsilon$ of dimensional regularization. On the contrary, evaluating a string correlator in order to model string scattering, the result will be again a double-expansion: the topological expansion parametrized by the string coupling $g_S$ and the expansion in the inverse string tension $\alpha'$.
It remains to be explored throughout the next years, whether those two double expansions can be related. Clearly, individual Feynman diagrams lead to divergent integrals, whose divergences cancel in the final, physical result only. This not being the case for string amplitudes points into the direction of a singular transformation. However, the idea of interpreting/identifying ,,stringyness'' simply as a regulating mechanism, which comes across very naturally, is rather appealing. 

\begin{acknowledgement}
Both authors would like to thank the Kolleg Mathematik und Physik Berlin for supporting the conference ``Antidifferentiation and the Calculation of Feynman Amplitudes''. AK would like to thank the IMPRS for Mathematical and Physical Aspects of Gravitation, Cosmology and Quantum Field Theory, of which he is a member and which renders his studies possible. 
\end{acknowledgement}
%



\end{document}